\documentclass[%
 reprint,
 amsmath,amssymb,
 aps, prd
]{revtex4-2}
\usepackage{graphicx}
\usepackage{bm}
\usepackage{xcolor}
\usepackage{hyperref}

\begin{document}

\title{High-cadence GW astrometry in the $\mu$Hz gap}

\author{Giorgio Mentasti}
\email{giorgio@apc.in2p3.fr}
\affiliation{Universit\'e Paris Cit\'e, CNRS, Astroparticule et Cosmologie, F-75013 Paris, France}

\author{Carlo~R.~Contaldi}
\affiliation{Department of Physics, Imperial College London, SW7 2AZ, London, United Kingdom}
\email{c.contaldi@imperial.ac.uk}

\date{\today}

\begin{abstract}
We forecast the sensitivity of a dedicated, fast-spinning astrometric survey to a
stochastic gravitational-wave background (SGWB) in the microhertz gap between pulsar
timing arrays (PTAs) and LISA. Spinning the satellite faster samples each star more
often, raising the Nyquist frequency and opening higher frequencies at no cost in noise: the
shorter, noisier exposures are exactly compensated by their number, so the white-noise
floor of each position stream is independent of cadence. High frequencies are where
astrometry is most competitive, because the deflection measures the strain directly: its
strain noise is white, whereas the equivalent noise of pulsar timing grows as $f^2$. With a
cross-correlation estimator over $N$ reference stars and the astrometric overlap computed in
the distant-source limit, we find that a Gaia-like satellite scanning the ecliptic poles,
with $N\sim10^4$ bright stars over $10$~yr, falls short of the background measured by PTAs
by a factor $\sim\!60$. A larger, $8$-m-class aperture, or a constellation of satellites
covering the whole sky, reaches it, while extending the reach continuously across the
$\mu$Hz band.
\end{abstract}

\maketitle

\section{Introduction}
\label{sec:intro}

Gravitational waves (GWs) are now observed across widely separated frequency
bands. Ground-based interferometers (LIGO, Virgo, KAGRA) routinely detect the
Hz--kHz signals of compact-binary mergers~\cite{LVKGWTC3}, while pulsar timing
arrays (PTAs), proposed as GW detectors decades ago~\cite{Detweiler1979}, have
recently reported evidence for a nanohertz stochastic GW
background (SGWB)~\cite{NANOGrav15yr,EPTADR2,PPTADR3,CPTADR1}, plausibly sourced by
the cosmic population of supermassive black-hole binaries (SMBHBs)~\cite{SesanaSMBHB2008}.
Beyond such astrophysical foregrounds, an SGWB may carry a cosmological component
from inflation, (p)reheating, first-order phase transitions or cosmic
strings~\cite{CapriniFigueroa2018,Christensen2019}, so that detecting it and
measuring its spectrum is a central goal of GW astronomy.

The space mission LISA will open the millihertz band~\cite{LISA2017}, but between
the PTA (nHz) and LISA (mHz) windows lies the \emph{microhertz gap}: several decades
of frequency that no current or planned observatory covers~\cite{SesanaMicroHz2021}.
The band is astrophysically rich---it hosts the inspiral of the same SMBHBs seen by
PTAs, here more massive and closer to coalescence, alongside cosmological backgrounds
whose energy density can rise toward higher frequency---yet it is notoriously hard to
reach, demanding either impractically long space-based baselines or impractically
high-cadence pulsar timing. Several ideas have been put forward to bridge it,
including precision ranging of inner solar-system asteroids used as test
masses~\cite{Fedderke2022}, extremely precise interferometric astrometry of a handful
of stable stars~\cite{FedderkeStellar2022}, resonant perturbations of binary systems monitored by
lunar/satellite laser ranging and binary pulsars~\cite{BlasJenkins2022} --- a route
recently forecast in detail for the SKA Observatory, where high-precision timing of a
handful of known binary millisecond pulsars is projected to reach the $\mu$Hz band and
to be complementary to conventional PTA analyses~\cite{JenkinsBlasFoster2026} --- and
high-cadence relative astrometry with photometric surveys and the Roman Space
Telescope~\cite{PardoRoman2023,Wang2021,Wang2022}.

Astrometry is a particularly natural probe of this band. A passing GW perturbs null
geodesics and deflects the apparent direction of a distant source by an angle of
order the strain itself, $\delta n\sim h$; an SGWB therefore imprints correlated
deflections on the sky, with an angular pattern---the astrometric analogue of the
Hellings--Downs curve~\cite{HellingsDowns1983}---built from vector spherical
harmonics~\cite{Pyne1996,BookFlanagan2011}. Crucially, the astrometric observable is
a \emph{direct} measure of the metric perturbation $h$: unlike the pulsar-timing
residual, which is the time integral of the GW-induced redshift and scales as
$\sim h/f$~\cite{Qin2019,Vaglio2025}, the deflection carries no factor of $1/f$. A
white measurement noise then yields a strain sensitivity that is \emph{flat} in
frequency for astrometry, whereas for timing the equivalent strain-noise power grows
as $f^2$; astrometry thus loses no sensitivity toward high frequencies and is
comparatively advantageous precisely at the upper, $\mu$Hz, end of its band. The
effect and its overlap functions have been developed in
detail~\cite{Mihaylov2018,Inomata2024,Caliskan2024,GolatContaldi2022,Perna2026},
including its sensitivity to parity violation~\cite{Liang2024,Jaraba2025}, and SGWB
searches have been forecast and performed with
Gaia~\cite{Klioner2018,Darling2018,Jaraba2023,Akhmetov2026,Geyer2025,Darling2025,Caldarola2026},
proposed for dedicated missions such as Theia~\cite{Theia2017}, and explored in
synergy with PTAs~\cite{Inomata2024,Vaglio2025,MentastiContaldi2024,FinkContaldiMentasti2025,JimenezCruz2025,PernaSKA2026,Filipello2026}.
Astrometric deflection is moreover only one of several ways the same metric perturbation
imprints itself on resolved images: a GW background also produces a time-dependent shear of
extended sources --- ``cosmic shimmering'' in time-resolved weak-lensing
data~\cite{Shimmering2024} --- and pulsar timing, astrometric deflection and shimmering can
be treated as one family of spin-weighted observables with closed-form auto- and
cross-correlations~\cite{SpinHarmonic2025}. The survey proposed here is a way to open the
$\mu$Hz decade for the astrometric member of that family.

Current surveys, however, are pinned to the low-frequency edge of this band by their
slow cadence: Gaia spins once every six hours, placing its nominal Nyquist frequency at
tens of $\mu$Hz, and revisits a given source only in brief groups of transits separated
by weeks; and quasar proper-motion analyses constrain
only the long-wavelength ($f\lesssim1/T$) regime~\cite{Akhmetov2026}. Reaching
deeper into the $\mu$Hz gap requires high-cadence sampling, as recently emphasized
for relative astrometry with photometric surveys and with the Roman Space
Telescope~\cite{PardoRoman2023,Wang2021,Wang2022}, and now carried through to end-to-end Bayesian
forecasts for archival \emph{Kepler} and forthcoming Roman data~\cite{Zhang2025}.

In this work we forecast the SGWB sensitivity of a dedicated, fast-spinning
astrometric survey whose scan crosses both ecliptic poles on every revolution, so that
each star in two small antipodal polar caps is sampled once per revolution. We describe that geometry, and anchor its noise model to
Gaia~\cite{GaiaMission2016,Fabricius2016}, in Sec.~\ref{sec:noise}, where we also show
that spinning such an
instrument faster does not degrade the noise floor of each position stream: the individual
exposures are shorter and therefore noisier, but there are proportionally more of them, and
for a white noise floor the two effects cancel exactly~\cite{Lindegren1978}. Cadence is
thus observable bandwidth bought for free.
We then build the optimal cross-correlation estimator (Sec.~\ref{sec:snr}), define the
survey configurations (Sec.~\ref{sec:configs}), and quote (Sec.~\ref{sec:numbers}) the threshold background amplitude as a function of source
number, aperture and spectral index. Throughout we follow the conventions of
Ref.~\cite{Vaglio2025}.

\section{Single-epoch astrometric noise}
\label{sec:noise}

\begin{figure}[t]
\centering
\includegraphics[width=0.93\columnwidth]{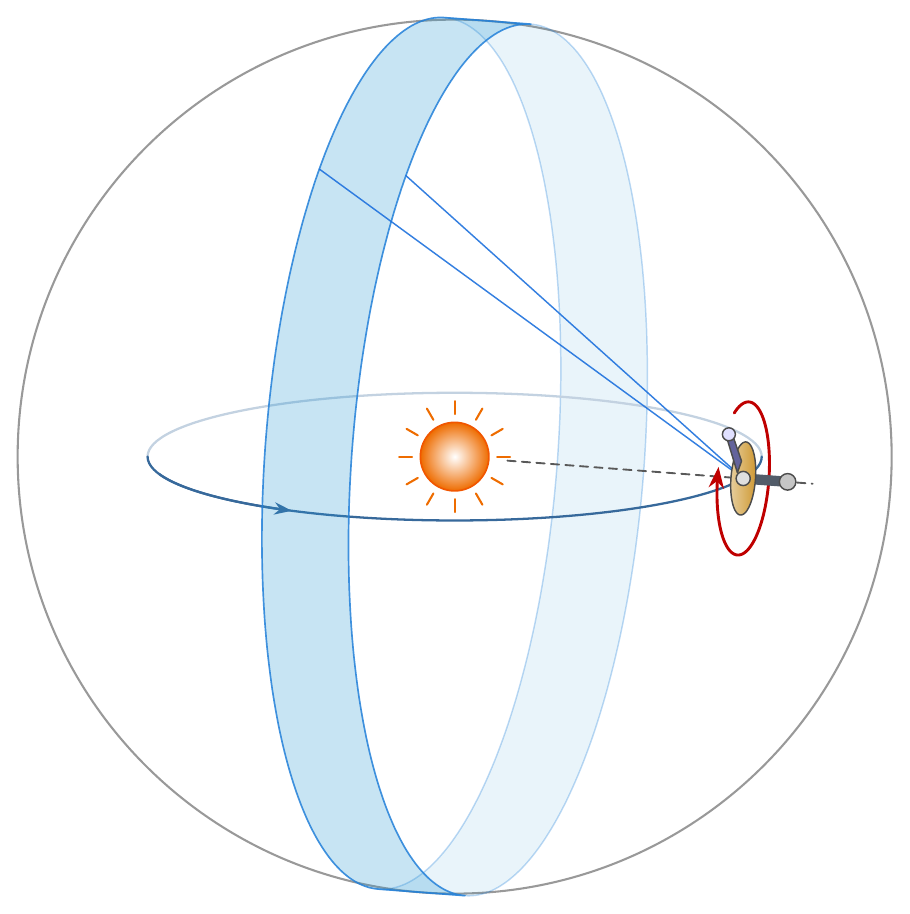}
\caption{Observing geometry (schematic, not to scale). The spacecraft (right) orbits the Sun
and spins about the Sun--satellite line (grey dashes; red arrow: one rotation). Its telescope
points perpendicular to that axis, so each rotation sweeps the field (blue lines) along the
shaded strip on the sky. The strip always passes through both ecliptic poles (top and
bottom), so the stars in two small polar caps are crossed on every revolution.
The figure was drawn as TikZ code by the AI model Claude Opus~5.5 (Anthropic) from a
hand sketch by the authors, who checked the rendered geometry against the sketch and the text.}
\label{fig:scheme}
\end{figure}

\subsection{The survey in one picture}
The instrument we have in mind is a single spacecraft on a circular heliocentric orbit,
spinning steadily about the Sun--satellite line, with its telescope pointing
\emph{perpendicular} to the spin axis (Fig.~\ref{fig:scheme}). Each revolution drags the
field around the great circle normal to the axis and paints the shaded strip across the
sky. This is Gaia's scanning architecture with one change. Gaia tilts its spin axis
$45^\circ$ away from the Sun and precesses it, so that a given star crosses the focal plane
in seconds, may be revisited a few hours later by the next revolution or by the second
field of view, and then does not return for weeks. Holding the axis on the Sun direction
instead keeps it in the ecliptic plane, and every great circle normal to such an axis passes
through both ecliptic poles. The strip therefore crosses both poles on every revolution, all
year: its orientation turns about the poles once a year as the Sun direction moves, but two
antipodal caps, of radius equal to the half-width of the field across the scan, never leave
it --- for the $5^\circ$ caps adopted below, a field about $10^\circ$ wide across the scan.
Every star in them is measured once per revolution, on a strictly uniform grid, for the
whole mission. Two consequences set up everything that follows: the sampling interval is the
spin period, so the accessible band extends up to $f_s/2$ and is bought simply by
spinning faster; and the references are a fixed, tightly clustered set rather than a
sparsely revisited all-sky catalogue, which is what makes their GW-induced deflections
correlate. The first is quantified in this section, the second in Sec.~\ref{sec:snr}.

Every quantity below is eventually evaluated against a real instrument, so we fix that
anchor at the outset. We calibrate the single-epoch error to Gaia, the only mission for
which photon-limited astrometry at this precision has actually been demonstrated in
flight. Its along-scan per-CCD centroiding is $\simeq0.06$~mas for a well-exposed bright
source at $G=12$ \cite{Fabricius2016,GaiaMission2016}; a field-of-view transit crosses
nine such CCDs, and averaging them gives $0.06/\sqrt9\simeq0.02$~mas per transit at that
magnitude. Photon noise then fixes the magnitude dependence: one magnitude is a factor
$10^{0.4}$ in flux and hence $10^{0.2}$ in centroiding error, so
$\sigma_c\propto10^{0.2G}$. Throughout this paper we adopt a \emph{single} fiducial
reference magnitude, $G=13$, and pivot the scaling there once and for all,
\begin{equation}
  \sigma_c(G) \simeq 0.032\,\text{mas}\times10^{0.2(G-13)},
  \qquad
  \sigma_c(13)\simeq0.032\,\text{mas},
  \label{eq:sigmaG}
\end{equation}
so that every number quoted in this paper refers to $G=13$ unless stated otherwise. What
remains is to establish how this per-epoch error responds to the spin rate, and what it
implies for the noise floor of the resulting time series.

\subsection{Photon-limited centroiding}
The fundamental per-epoch (single field-of-view transit) astrometric error is set
by photon-limited centroiding of the point-spread function (PSF)~\cite{Lindegren1978},
\begin{equation}
  \sigma_c \;\simeq\; \frac{\sigma_{\rm PSF}}{\sqrt{N_{\rm ph}}},
  \qquad
  \sigma_{\rm PSF}\;\simeq\;\frac{\lambda}{D},
  \label{eq:centroid}
\end{equation}
where $\sigma_{\rm PSF}\simeq\lambda/D$ is the angular width of the
diffraction-limited PSF ($D$ the along-scan aperture, $\lambda$ the effective
wavelength) and $N_{\rm ph}$ the number of detected photons per transit.
Equation~\eqref{eq:centroid} is the photon-noise (Cram\'er--Rao) centroiding limit
in which the centroid of a source of $N_{\rm ph}$ photons spread over a width
$\sigma_{\rm PSF}$ is located to a precision $\sigma_{\rm PSF}/\sqrt{N_{\rm ph}}$
\cite{Lindegren1978}; for Gaia it is realised at the sub-mas level per CCD
\cite{Fabricius2016,GaiaMission2016}.
The number of detected photons per transit is
\begin{equation}
  N_{\rm ph} \;=\; \Phi_\star \, t_{\rm int},
  \label{eq:Nph}
\end{equation}
where $\Phi_\star$ (photons s$^{-1}$) is fixed by aperture and source flux.
For a survey that spins at frequency $f_s$ and lets each source cross the focal
plane once per revolution, the integration time per transit scales as
\begin{equation}
  t_{\rm int} \;=\; \frac{\eta}{f_s}
  \label{eq:tint}
\end{equation}
with $\eta$ the fraction of each revolution the source spends crossing the focal plane,
i.e.\ the along-scan width of the field over $360^\circ$: $\eta\simeq2\times10^{-3}$ for
Gaia's $\simeq0.7^\circ$ astrometric field, and the same for the Gaia-class instrument
considered here. Combining
\eqref{eq:centroid}--\eqref{eq:tint},
\begin{equation}
  \boxed{\;
  \sigma_c(f_s) \;=\; \sigma_{\rm PSF}\,
  \sqrt{\frac{f_s}{\eta\,\Phi_\star}}
  \;\equiv\; \sigma_0\sqrt{\frac{f_s}{f_0}}\; }
  \label{eq:sigma_fs}
\end{equation}
i.e. the per-epoch error grows as $\sqrt{f_s}$. The second form is the practical one:
$\sigma_0\equiv\sigma_{\rm PSF}\sqrt{f_0/\eta\Phi_\star}$ is the per-epoch error measured
at some reference spin rate $f_0$, so a single-epoch precision is only meaningful when
quoted together with the cadence at which it was achieved. This is exactly how the Gaia anchor of
Eq.~\eqref{eq:sigmaG} is used below, with
$(\sigma_0,f_0)=(\sigma_c(G),f_s^{\rm ref})$ [Eq.~\eqref{eq:SigmaG}].
The spin frequency also fixes the sampling and the Nyquist edge of the band,
\begin{equation}
  \Delta t = \frac{1}{f_s},
  \qquad
  f_{\rm Ny} = \frac{f_s}{2}.
  \label{eq:nyquist}
\end{equation}
The white-noise power spectral density of each source position stream is then, written
as a one-sided PSD (frequencies $f>0$ only), as in Eq.~(21) of Ref.~\cite{Vaglio2025},
\begin{equation}
  \boxed{\;
  \Sigma \;=\; 2\,\sigma_c^2\,\Delta t
  \;=\; \frac{2\,\sigma_{\rm PSF}^2}{\eta\,\Phi_\star}
  \;=\; \text{const}\;}
  \label{eq:Sigma}
\end{equation}
The growth $\sigma_c^2\propto f_s$ cancels the cadence gain $\Delta t\propto
1/f_s$: the noise \emph{floor} $\Sigma$ is set by aperture and source flux alone,
while $f_s$ sets only the observable \emph{bandwidth} $[1/T,\,f_{\rm Ny}]$.

Physically, a star delivers a fixed photon budget per unit time: spinning faster only slices
it into more, shorter, noisier exposures, and since a white noise floor is the variance per
sample times the time between samples, the two effects cancel. The extra samples buy Nyquist
range at the same noise level --- cadence is free bandwidth. This is specific to astrometry:
for pulsar timing the strain-referred noise carries an extra factor $f^2$, so the
high-frequency end is suppressed however often one samples.

Because $\sigma_c$ and $\Delta t$ are coupled, the Gaia anchor of Eq.~\eqref{eq:sigmaG},
measured at the Gaia spin rate $f_s^{\rm ref}=1/6\,\text{h}=4.6\times10^{-5}$~Hz, must be
paired with the Gaia sampling interval,
\begin{equation}
  \Sigma(G) = \frac{2\,\sigma_c(G)^2}{f_s^{\rm ref}}
  \;=\; \frac{2\,\sigma_0^2}{f_0},
  \qquad
  \Sigma(13)\simeq 1.0\times10^{-15}\,\text{s}.
  \label{eq:SigmaG}
\end{equation}
This is a calibration, not a residual cadence dependence: spinning faster leaves $\Sigma$
unchanged [Eq.~\eqref{eq:Sigma}]. Equation~\eqref{eq:SigmaG} is the noise floor adopted in
every estimate below.

The white-noise reading needs a uniformly sampled stream. A Gaia-like scanning law returns
sparse, uneven visits whose gaps alias power across the spectrum and leave no clean Nyquist
edge. Holding the spin axis on the Sun removes the gaps for the polar caps: the same stars
are crossed on a uniform grid $\Delta t=1/f_s$ for the whole mission, and with the Sun always
$90^\circ$ from the telescope no re-pointing or seasonal interruption is needed over
$T=10$~yr.

We take bright stars ($G\simeq13$) as references: the all-sky supply is already
$\sim\!10^{7}$ sources~\cite{GaiaDR3}, and centroiding is still photon-limited
[Eq.~\eqref{eq:sigmaG}]. Their parallax and proper motion are deterministic, confined to
$f\lesssim1/\text{yr}$, and removed by a five-parameter astrometric fit per star at the cost
of the lowest few frequency bins (Sec.~\ref{sec:numbers}). The magnitude cut is not
critical: going fainter raises the noise floor as $\Sigma\propto10^{0.4G}$ but the star
counts as $N(\!<\!G)\propto10^{0.35G}$, and the two nearly cancel in $\Omega_0\propto\Sigma/N$
(Sec.~\ref{sec:budget}).

\section{SGWB signal and signal-to-noise}
\label{sec:snr}

\subsection{Setup}
A passing GW shifts the apparent direction $\mathbf n\to\mathbf n+\delta\mathbf n$.
For each source $I=1,\dots,N$ we project the deflection onto two orthogonal
angular components $a\in\{\theta,\phi\}$ and form the time streams
\begin{equation}
  d_I^{a}(t) \;=\; \delta n_I^{a}(t) + \zeta_I^{a}(t),
  \label{eq:stream}
\end{equation}
with $\zeta_I^a$ the measurement noise.
Assuming the noise is white, Gaussian, stationary and uncorrelated between sources
and components,
\begin{equation}
  \big\langle \tilde\zeta_I^{a}(f)\,\tilde\zeta_J^{b\star}(f')\big\rangle
  = \delta_{IJ}\,\delta^{ab}\,\delta(f-f')\,\Sigma,
  \qquad \Sigma=2\,\sigma_c^2\Delta t.
  \label{eq:noisePSD}
\end{equation}
Equations~\eqref{eq:noisePSD} and \eqref{eq:signalPSD} are written in the
\emph{two-sided} convention of Ref.~\cite{Vaglio2025}, while the frequency integrals of Eqs.~\eqref{eq:snr_opt} and
\eqref{eq:snr_integral} then fold the negative frequencies onto $f>0$, which is the origin
of the factor $2$ in $\Sigma=2\sigma_c^2\Delta t$ and of the Allen--Romano $2T$ quoted
below. Because signal and noise are referred to the same convention throughout, every SNR
and every threshold in this paper depends only on the ratio $S_h/\Sigma$ and is therefore
independent of the choice. For a stationary, isotropic, Gaussian SGWB the signal
cross-spectrum carries the angular pattern
\begin{equation}
  \big\langle \delta \tilde{n}_I^{a}(f)\,
  \delta \tilde{n}_J^{b\star}(f')\big\rangle
  = \delta(f-f')\,S_h(f)\,\Gamma^{ab}(\hat{\mathbf n}_I,\hat{\mathbf n}_J),
  \label{eq:signalPSD}
\end{equation}
where $S_h(f)$ is the one-sided strain PSD [Eq.~\eqref{eq:omega} below] and
$\Gamma^{ab}(\hat{\mathbf n}_I,\hat{\mathbf n}_J)$ is the astrometric overlap between
the $a$ component of the first star and the $b$ component of the second, a tensor in both
directions [Eq.~(44) of Ref.~\cite{Mihaylov2018}]. Our stars lie far beyond the GW
wavelength, in the \emph{distant-source} limit where only the ``Earth term''
survives~\cite{BookFlanagan2011,Mihaylov2018,Inomata2024}. There the response of a single
source reduces to its total mean-square deflection, Eq.~(3) of Ref.~\cite{BookFlanagan2011},
\begin{equation}
  \big\langle|\delta\mathbf n|^2\big\rangle
  =\frac{1}{4\pi^2}\int\! d\ln f\;\Big(\frac{H_0}{f}\Big)^{2}\Omega_{\rm GW}(f),
  \label{eq:bf3}
\end{equation}
and inserting $\Omega_{\rm GW}=(2\pi^2/3H_0^2)f^2h_c^2$ with $S_h=h_c^2/f$
[Eq.~\eqref{eq:omega} below] collapses the prefactor to $1/6$:
\begin{equation}
  \big\langle|\delta\mathbf n|^2\big\rangle=\frac16\int\! S_h(f)\,df .
  \label{eq:dnvar}
\end{equation}
How that variance divides between the two components is fixed by the tensor structure of
their two-point function, Eq.~(64) of Ref.~\cite{BookFlanagan2011}, whose angular factor
reduces at coincidence to the transverse projector,
$H_{ij}(\mathbf n,\mathbf n)\propto\delta_{ij}-n_in_j$ [their Eq.~(83)]. In the orthonormal
basis $(\hat{\boldsymbol\theta},\hat{\boldsymbol\phi})$ that projector is the
$2\times2$ identity, so the two transverse components carry equal variance and are
mutually uncorrelated. Splitting Eq.~\eqref{eq:dnvar} accordingly and comparing with the
definition $\langle\delta n^a\delta n^{b\star}\rangle=S_h\Gamma^{ab}$ of
Eq.~\eqref{eq:signalPSD} gives
\begin{equation}
  \Gamma^{\theta\theta}(\hat{\mathbf n},\hat{\mathbf n})
  =\Gamma^{\phi\phi}(\hat{\mathbf n},\hat{\mathbf n})=\tfrac1{12},
  \qquad \Gamma^{\theta\phi}(\hat{\mathbf n},\hat{\mathbf n})=0,
  \label{eq:overlap0}
\end{equation}
the vanishing of the cross term being the statement that the projector is diagonal in this
basis, equivalently that parity forbids it for an unpolarised
background~\cite{Mihaylov2018,Caliskan2024}. At coincidence the overlap is $\tfrac1{12}$
times the identity, so Eq.~\eqref{eq:overlap0} holds in any orthonormal tangent basis; for
separated stars the off-diagonal components in a fixed basis are in general nonzero.

\subsection{Estimator and SNR}
The logic is the one standard for stochastic signals. A single star's stream cannot be
distinguished from its own noise, because the background is itself noise-like and adds
no feature to a single power spectrum. The GW-induced deflections of two
\emph{different} stars, however, are correlated with each other, while their
measurement noises are not. Cross-correlating a pair isolates that common component,
and the correlation it must have as a function of the angle between the two stars ---
the astrometric analogue of the Hellings--Downs curve --- is what turns an excess into
a signature. Accordingly, for each pair $I\neq J$ and each pair of components $(a,b)$ we
build the optimally filtered cross-correlator
\begin{equation}
  \begin{gathered}
  \mathcal{C}_{IJ}^{ab}
  \equiv \int\! df\,\tilde d_I^{a}(f)\,\tilde d_J^{b\star}(f)\,\tilde Q(f),\\[2pt]
  \langle\mathcal{C}_{IJ}^{ab}\rangle
  = \!\int\! df\, S_h Q\,\Gamma^{ab}(\hat{\mathbf n}_I,\hat{\mathbf n}_J),
  \end{gathered}
  \label{eq:estimator}
\end{equation}
with expected signal-to-noise
\begin{equation}
  \mathrm{SNR}^2
  = \sum_{I\neq J}\sum_{a,b}
  \frac{\big|\langle\mathcal{C}_{IJ}^{ab}\rangle\big|^2}
       {\big\langle|\mathcal{C}_{IJ}^{ab}|^2\big\rangle},
  \label{eq:snr_def}
\end{equation}
and its optimal, matched-filter combination over all pairs and components, including the
GW contribution to each stream's auto-power, is~\cite{Vaglio2025,MentastiContaldi2024,FinkContaldiMentasti2025}
\begin{equation}
  \mathrm{SNR}^2_{\rm opt}
  \;\simeq\; \overline{\Gamma^2}\,N^2\,T
  \int_{1/T}^{f_{\rm Ny}} \! df\,
  \frac{S_h^2(f)}{\big[\Sigma+\overline{\Gamma}\,S_h(f)\big]^2},
  \label{eq:snr_opt}
\end{equation}
where $\overline{\Gamma^2}\equiv\big\langle\sum_{a,b}\big[\Gamma^{ab}(\hat{\mathbf n}_I,\hat{\mathbf n}_J)\big]^2\big\rangle_{\rm pairs}$
is the pair-averaged squared overlap, independent of the tangent basis chosen at each star,
and $\overline{\Gamma}\equiv(\overline{\Gamma^2})^{1/2}$. Equation~\eqref{eq:snr_opt} is valid in the general (not necessarily noise-dominated) regime, the analogue of
Eq.~(34) of Ref.~\cite{Vaglio2025}, to which it reduces for $\overline{\Gamma}\to1$, the
approximation made there. Carrying the response through both the numerator and the
denominator is what makes Eq.~\eqref{eq:snr_opt} well behaved at strong signal: as
$\Sigma\to0$ the two factors of $\overline{\Gamma}$ cancel and the SNR becomes independent
of the overlap, as it must once the measurement is a ratio of signal to signal. The survey
considered here sits in the opposite, deeply noise-dominated regime, where the denominator
is $\simeq\Sigma^2$ and the weight of the $S_h$ term is numerically immaterial (below).
The factors of two follow that reference: the
single-pair statistic carries the Allen--Romano factor $2T$ from folding the
negative frequencies, and summing it over the $\simeq N^2/2$
independent (unordered)
source pairs cancels that $2$ against the $1/2$, leaving the prefactor $N^2T$ with
the frequency integral over $f>0$ and $\overline{\Gamma^2}$ the pair-averaged squared
overlap [cf.~Eq.~(34) of Ref.~\cite{Vaglio2025}]. Following the same reference, we
parametrise the background by the characteristic strain,
\begin{equation}
  \begin{aligned}
  h_c(f)&=A_{\rm GW}\left(\frac{f}{f_{\rm ref}}\right)^{\alpha},\\[2pt]
  S_h(f)&=\frac{h_c^2(f)}{f}
         =B\left(\frac{f}{f_{\rm ref}}\right)^{2\alpha-1},\\[2pt]
  \Omega_{\rm GW}(f)&=\frac{2\pi^2}{3H_0^2}\,f^2 h_c^2(f),
  \end{aligned}
  \label{eq:omega}
\end{equation}
where $B\equiv3H_0^2\Omega_0/2\pi^2 f_{\rm ref}^3$, with $\alpha$ the characteristic-strain
spectral index, $\Omega_0\equiv\Omega_{\rm GW}(f_{\rm ref})$ the amplitude at the pivot
$f_{\rm ref}=30$~nHz, $S_h$ the one-sided strain PSD, and $H_0$ the Hubble constant; a circular-orbit SMBHB
background has $\alpha=-2/3$. The astrometric response is carried by the overlap
$\Gamma^{ab}$ in Eq.~\eqref{eq:signalPSD}~\cite{Mihaylov2018,Inomata2024,Vaglio2025}.
Inserting this power law, Eq.~\eqref{eq:snr_opt} becomes the working expression
\begin{equation}
  \mathrm{SNR}^2_{\rm opt}
  \simeq \overline{\Gamma^2}\,N^2\,T
  \int_{1/T}^{f_{\rm Ny}}\!\! df\,
  \frac{\big[B\,(f/f_{\rm ref})^{2\alpha-1}\big]^2}
       {\big[\Sigma+\overline{\Gamma}\,B\,(f/f_{\rm ref})^{2\alpha-1}\big]^2},
  \label{eq:snr_integral}
\end{equation}
which is nonlinear in $\Omega_0$ (through $B$) and is solved numerically for the
threshold $\Omega_0(\mathrm{SNR}=1)$ in the general regime; this is the expression
used for the estimates below. For the amplitudes of interest the survey turns out
deeply noise-dominated. The quantity that actually competes with $\Sigma$ in
Eq.~\eqref{eq:snr_opt} is $\overline{\Gamma}S_h$, and at the $\mathrm{SNR}=1$ threshold its
largest value anywhere in the band, over all configurations and the spectral indices of
Table~\ref{tab:thresholds}, is
$\overline{\Gamma}S_h/\Sigma\simeq2.2\times10^{-4}$ (at $\alpha=-1$;
the bare ratio is $S_h/\Sigma\simeq2.7\times10^{-3}$). The general and noise-dominated
thresholds therefore agree to $\lesssim0.02\%$, the denominator
is $\simeq\Sigma^2$, and the cadence dependence reduces
to $I(\alpha)=\int_{1/T}^{f_{\rm Ny}}(f/f_{\rm ref})^{2(2\alpha-1)}df$, which has two
regimes separated by $\alpha=1/4$. For $\alpha<1/4$ (red) it is dominated by the lower
edge $1/T$ and for $\alpha>1/4$ (blue) by the Nyquist edge $f_{\rm Ny}$, giving
\begin{align}
  \text{red}\ (\alpha<\tfrac14): &\quad
  \mathrm{SNR}\propto T^{\,1-2\alpha}\ \ (\text{cadence-independent}),
  \label{eq:red}\\[2pt]
  \text{blue}\ (\alpha>\tfrac14): &\quad
  \mathrm{SNR}^2\propto f_s^{\,4\alpha-1}\ \ (\text{grows with spin}).
  \label{eq:blue}
\end{align}
For red spectra [Eq.~\eqref{eq:red}] the SNR is set by total integration time and is
independent of cadence; for flat/blue spectra it grows with the Nyquist edge
$f_{\rm Ny}=f_s/2$, so a faster spin directly buys sensitivity.

We use the characteristic-strain index $\alpha$ ($h_c\propto f^{\alpha}$), with the
normalisation of Eq.~(29) of Ref.~\cite{Vaglio2025}: a circular-orbit SMBHB background has $\alpha=-2/3$, while a
scale-invariant energy density ($\Omega_{\rm GW}\propto f^0$) corresponds to
$\alpha=-1$ (since $\Omega_{\rm GW}\propto f^{2\alpha+2}$). Note that the often-quoted
$13/3$ is the PTA \emph{timing-residual} PSD index ($S_r\propto f^{-13/3}$, residual
$\propto h/f$), not the deflection spectrum $S_h\propto f^{2\alpha-1}$ used here. The two
indices are related by
\begin{equation}
  \gamma = 3-2\alpha,
  \label{eq:gamma}
\end{equation}
so the SMBHB value $\alpha=-2/3$ is $\gamma=13/3$; we label the rows of
Table~\ref{tab:thresholds} by both. For reference, the NANOGrav 15-yr posterior peaks near
$\gamma\simeq3.2$, i.e.\ $\alpha\simeq-0.1$~\cite{NANOGrav15yr}, which the
$\alpha=-1/3$ and $\alpha=0$ rows bracket.

\subsection{Broadband reach: the power-law integrated curve}
\begin{figure*}[t]
\centering
\includegraphics[width=0.48\textwidth]{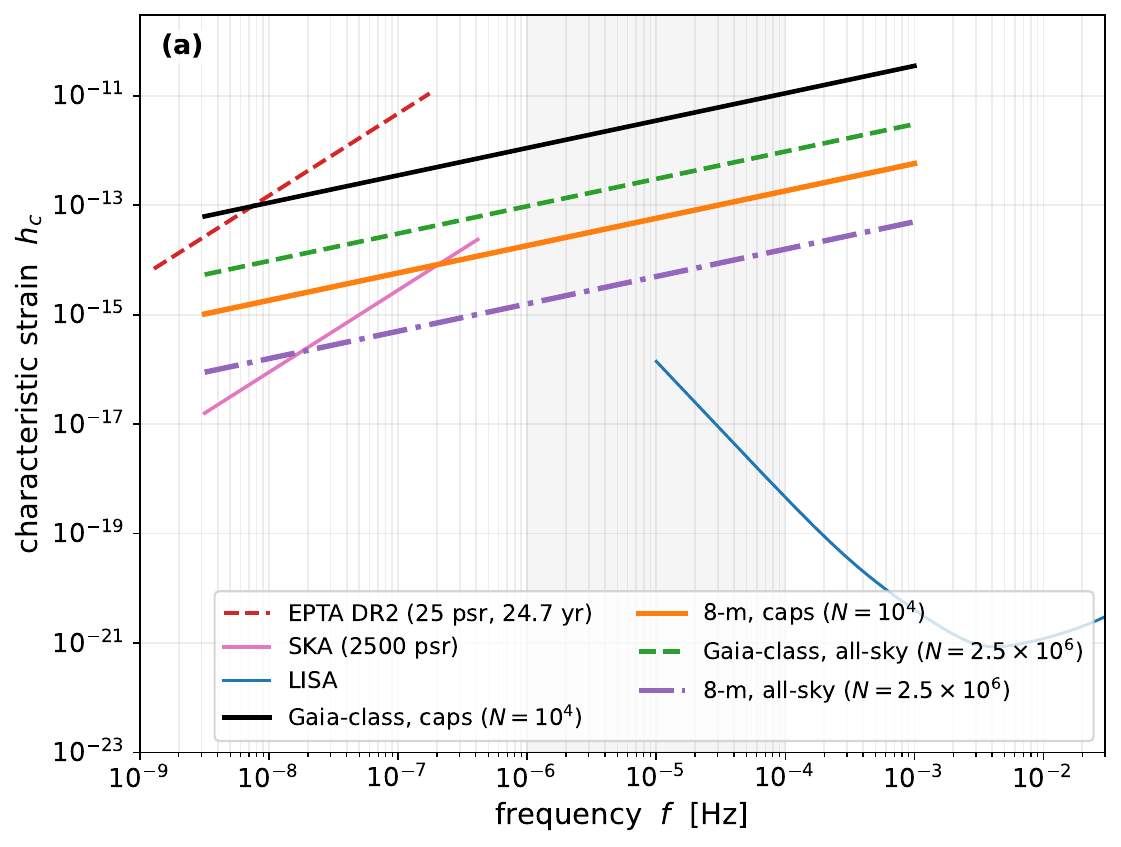}\hfill
\includegraphics[width=0.48\textwidth]{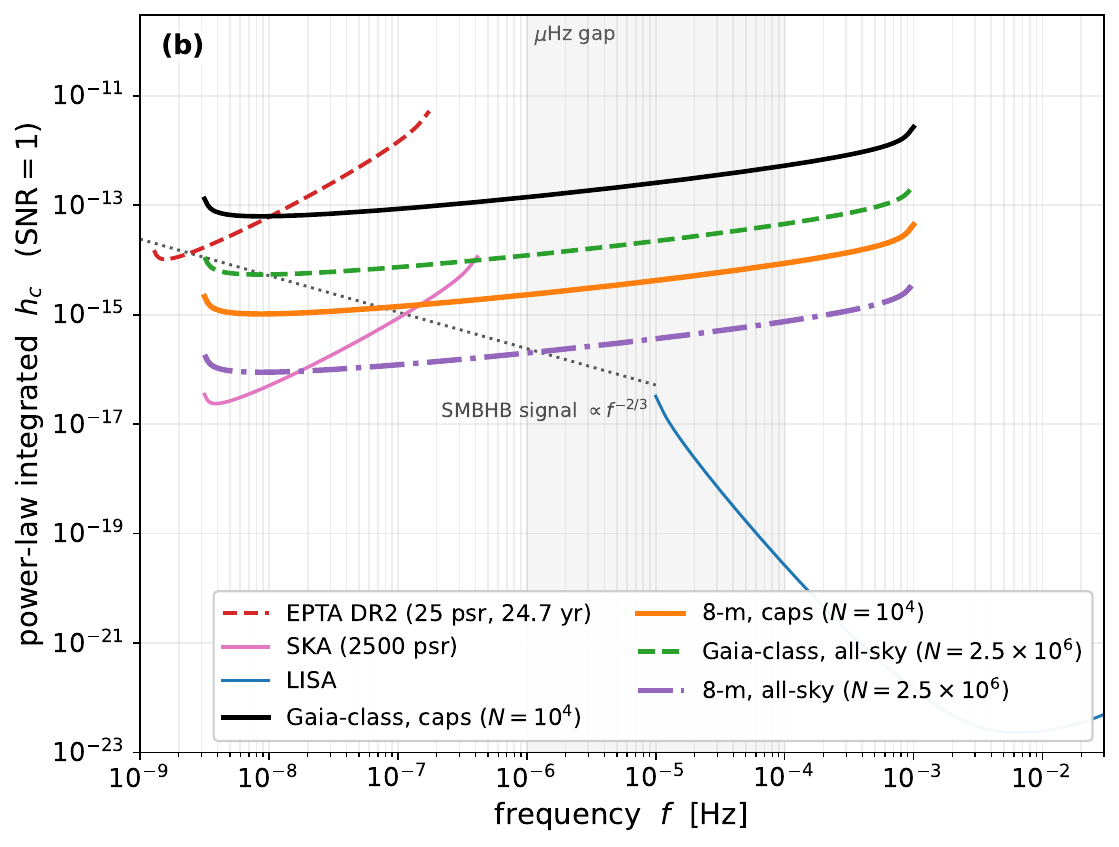}
\caption{Reach of the astrometric survey against PTAs and LISA ($\mu$Hz gap shaded), in
two views of the \emph{same}
effective noise $S_{\rm eff}(f)\equiv[h_c^{\rm sens}(f)]^2/f$. \textbf{(a)} Per-bin
($\Delta f=1/T_{\rm obs}$) $\mathrm{SNR}=1$ characteristic strain, Eq.~\eqref{eq:hcsens},
rising as $\sqrt f$ for the survey because $\Sigma$ is white. \textbf{(b)} Power-law
integrated sensitivity~\cite{ThraneRomano2013,FinkContaldiMentasti2025} in the same
variable, Eq.~\eqref{eq:pls}, on the same axes: a power law \emph{tangent} to a curve is
detected at $\mathrm{SNR}=1$; dotted, the measured PTA background
$h_c=2.4\times10^{-15}(f/f_{\rm yr})^{-2/3}$, drawn to $10\,\mu$Hz. The four survey curves
form a $2\times2$ in aperture and sky coverage at one magnitude $G=13$ and share
$T_{\rm obs}=10$~yr with SKA and LISA; EPTA is drawn on its real $24.7$-yr DR2 baseline. The
PTA curves are deliberately simple toy models. Conventions, parameters and caveats in
Sec.~\ref{sec:numbers}. Both panels were produced by Python code written by the AI
model Claude Opus~5.5 (Anthropic) under the authors' direction and checked by them as
described in Sec.~\ref{sec:numbers}.}
\label{fig:sens}
\end{figure*}

Evaluated in a single bin of width $\Delta f=1/T$, Eq.~\eqref{eq:snr_opt} answers ``how loud
must the background be in \emph{this} bin''. A real SGWB is broadband and the estimator sums
every bin, so a power law can be detected even when it lies above that per-bin answer
everywhere: a per-bin statement understates the reach, by an amount depending on the
spectral index. It is convenient first to collapse the array into one effective strain
noise,
\begin{equation}
  S_{\rm eff}(f)\equiv\frac{S_n(f)}{\overline{\Gamma}\,N_{\rm src}},
  \qquad
  \mathrm{SNR}^2=T\!\!\int_{f_{\rm lo}}^{f_{\rm hi}}\!\!\! df
  \left[\frac{S_h(f)}{S_{\rm eff}(f)}\right]^{2},
  \label{eq:seff}
\end{equation}
with $S_n$ the strain-referred noise power of a single source and $N_{\rm src}$ the number
of them cross-correlated; for the survey $S_n=\Sigma$ and $N_{\rm src}=N$, and
Eq.~\eqref{eq:seff} is exactly the noise-dominated limit of Eq.~\eqref{eq:snr_opt}. Written
this way the same expression serves any cross-correlation experiment, which is what
Sec.~\ref{sec:numbers} exploits to put PTAs on the same footing. The index-independent
statement one then wants is the power-law integrated sensitivity
(PLS)~\cite{ThraneRomano2013}, which we quote in characteristic strain, as in
Ref.~\cite{FinkContaldiMentasti2025}. For each index $\alpha$ one writes
$h_c=A_\alpha(f/f_{\rm ref})^\alpha$, solves $\mathrm{SNR}=1$ for its amplitude $A_\alpha$
and plots the envelope of all of them; with $S_h=h_c^2/f$ this is
\begin{equation}
  \begin{gathered}
  A_\alpha=\Bigg[T\!\!\int_{f_{\rm lo}}^{f_{\rm hi}}\!\!\!df
  \left(\frac{(f/f_{\rm ref})^{2\alpha}}{f\,S_{\rm eff}(f)}\right)^{\!2}\,\Bigg]^{-1/4},\\[4pt]
  h_c^{\rm PLS}(f)=\max_{\alpha}\ A_\alpha\left(\frac{f}{f_{\rm ref}}\right)^{\!\alpha}.
  \end{gathered}
  \label{eq:pls}
\end{equation}
The quarter power is simply because $h_c$ enters $\mathrm{SNR}$ squared. Since
$h_c\propto f^{-1}\Omega_{\rm GW}^{1/2}$ is monotonic in $\Omega_{\rm GW}$ at fixed $f$, the
envelope commutes with the change of variable: Eq.~\eqref{eq:pls} is the same curve as the
usual $\Omega_{\rm PLS}$, read in strain. It has the property one wants from a sensitivity
plot: a power law tangent to it is detected at exactly $\mathrm{SNR}=1$, and one rising
anywhere above it is detected with more, so ``touching the curve'' means a detection. Both
curves are drawn in Fig.~\ref{fig:sens} and read off in Sec.~\ref{sec:numbers}.

\begin{table}[!t]
\caption{Threshold $\Omega_0$ at $\mathrm{SNR}=1$ (Gaia-class, $G=13$, $T=10$~yr,
$\overline{\Gamma^2}=6.8\times10^{-3}$, pivot $f_{\rm ref}=30$~nHz), from the general
Eq.~\eqref{eq:snr_integral}, for a realistic $N=10^4$ bright references
($\Omega_0\propto[\sqrt{\overline{\Gamma^2}}\,N]^{-1}$). Red rows ($\alpha<1/4$) are
spin-independent; the blue row ($\alpha=1$) is quoted at the realistic Nyquist edge
$f_{\rm Ny}=1$~mHz (see text). Rows are labelled by both the characteristic-strain index
$\alpha$ and the equivalent PTA timing index $\gamma=3-2\alpha$ [Eq.~\eqref{eq:gamma}];
the energy density runs as $\Omega_{\rm GW}\propto f^{2\alpha+2}$, and the NANOGrav 15-yr
posterior, $\gamma\simeq3.2$ ($\alpha\simeq-0.1$), falls between the third and fourth rows.
$\Omega_0$ is \emph{non-monotonic} in $\alpha$, peaking between $\alpha=0$ and the
red/blue transition $\alpha=1/4$ --- a pivot artifact (see text).}
\begin{ruledtabular}
\begin{tabular}{lc}
 $\alpha$ & $\Omega_0$ ($N=10^4$) \\
\colrule
 $-1$ \ (scale-inv., $\gamma=5$)   & $1.1\times10^{-7}$ \\
 $-2/3$ \ (SMBHB, $\gamma=13/3$)   & $4.3\times10^{-7}$ \\
 $-1/3$ \ ($\gamma=11/3$)          & $1.5\times10^{-6}$ \\
 $0$ \ ($\gamma=3$)                & $4.5\times10^{-6}$ \\
 $1$ \ (blue, $\gamma=1$)          & $4.0\times10^{-12}$ \\
\end{tabular}
\end{ruledtabular}
\label{tab:thresholds}
\end{table}

\section{Survey configurations}
\label{sec:configs}

We consider four survey configurations, a $2\times2$ in aperture (Gaia-class or $8$-m) and in
sky coverage (two polar caps or the whole sky), all at the single magnitude $G=13$. This
section fixes their source counts and overlaps; Sec.~\ref{sec:numbers} gives their reach.

\subsection{Polar caps and the overlap}
\label{sec:caps}
In the frame adapted to each pair --- deflection components parallel and perpendicular to the
great circle joining the two stars [Eq.~(44) of Ref.~\cite{Mihaylov2018}] --- the overlap is
$\Gamma(\Theta)$ times the identity, with $\Gamma(\Theta)=\tfrac1{12}T(\Theta)$ and $T$,
normalised to $T(0)=1$, the astrometric analogue of the Hellings--Downs curve [their
Eq.~(45)], now measured on Gaia quasar proper motions~\cite{Darling2025}. It is maximal at
coincidence, crosses zero near $57^\circ$, reaches $-0.44$ around $100^\circ$, and vanishes
for antipodal stars. Clustering the references in small caps therefore keeps every
within-cap pair near the maximum, while pairs with one star in each cap carry no signal:
the optimal estimator reduces to two independent within-cap cross-correlations combined in
quadrature, and the second cap contributes by doubling $N$.

Quantitatively, at coincidence Eq.~\eqref{eq:overlap0} gives
$\overline{\Gamma^2}=2\,(1/12)^2=1/72\approx1.4\times10^{-2}$, the maximum attainable.
Averaging $\sum_{a,b}(\Gamma^{ab})^2=2\,\Gamma(\Theta)^2$ over pairs drawn uniformly from two
antipodal $5^\circ$ caps, with half the stars in each, gives $\overline{\Gamma^2}\simeq6.8\times10^{-3}$: the within-cap pairs retain $1.35\times10^{-2}$, $97\%$ of the ceiling, while the cross-cap
pairs contribute $\sim\!3\times10^{-7}$, so the average is very nearly half the ceiling. Since
$\Omega_0\propto1/\sqrt{\overline{\Gamma^2}}$, this still buys a factor $\simeq1.9$ in reach
over the same stars spread across the whole sky [Eq.~\eqref{eq:gbarsky}]. A caveat
attaches to the same feature and is taken up with the other systematics below:
nearly coincident pairs are the configuration a star-referenced attitude solution is
least able to preserve.

\subsection{Source budget and choice of targets}
\label{sec:budget}
Two $5^\circ$ caps cover a sky fraction $1-\cos\theta_{\rm cap}\approx4\times10^{-3}$, i.e.\
$\sim\!4\times10^4$ of the $\sim\!10^7$ stars with $G\lesssim13$; excluding variables and
resolved binaries leaves $N\sim10^4$, while full-sky coverage
(Sec.~\ref{sec:constellation}) gives $N=2.5\times10^6$, a factor $250$. The magnitude is not
a lever: with $\Sigma\propto10^{0.4G}$ and $N(\!<\!G)\propto10^{0.35G}$,
$\Omega_0\propto\Sigma/N\propto10^{0.05G}$, so depth is neutral and every configuration is
held at $G=13$; the only levers are aperture ($\Sigma\propto\lambda^2/D^4$) and sky coverage
($N$) [Eq.~\eqref{eq:scaling}]. We use stars rather than quasars: the absence of proper
motion that makes quasars attractive is moot here, since stellar motions are removed by the
astrometric fit (Sec.~\ref{sec:noise}), while at $G\sim19$ quasars would raise $\Sigma$, and
$\Omega_0$, by $10^{0.4\times6}\approx250$ (the $\alpha=-1$ entry of
Table~\ref{tab:thresholds} would rise from $1.1\times10^{-7}$ to $2.8\times10^{-5}$).
Throughout, $T=10$~yr, $f_{\rm ref}=30$~nHz, and $\Omega_0$ is the $\mathrm{SNR}=1$
threshold of Eq.~\eqref{eq:snr_integral}.

\subsection{All-sky constellation}
\label{sec:constellation}
A single satellite monitors only the two ecliptic polar caps. A \emph{constellation} of $k$
such satellites, distributed in longitude along the Earth's heliocentric orbit, extends the
coverage to the whole sky at the same cadence: each scans the great circle through the
ecliptic poles normal to its own Sun direction, so satellites spaced by $\Delta\varphi$ in
orbital longitude scan circles rotated by $\Delta\varphi$, whereas satellites near the Earth
would all scan the same circle. Covering the ecliptic equator without gaps, with strips of
half-width $\theta_{\rm cap}$, requires
\begin{equation}
  k \;\gtrsim\; \frac{\pi}{2\,\theta_{\rm cap}}\;\simeq\;18
  \qquad(\theta_{\rm cap}=5^\circ),
  \label{eq:ksat}
\end{equation}
spread over $\pm90^\circ$ of orbital longitude about the Earth. Away from the poles each
stream is stitched from successive satellites, every $\simeq10$~days at the equator, which
requires synchronised spin phases and makes inter-satellite calibration a systematic
specific to this configuration. The reference budget then grows to the full-sky
count at the \emph{same} fiducial magnitude: of the $\sim\!10^7$ sources with
$G\!\lesssim\!13$, the cleanliness cut of Sec.~\ref{sec:budget} leaves
$N\!\sim\!2.5\times10^{6}$ astrometrically stable references --- some $250$ times the
single-cap $N=10^4$, and the hard ceiling on any full-sky survey that stays at $G=13$.

There is a price for going full-sky, and it falls on the overlap. The
$\overline{\Gamma^2}\simeq6.8\times10^{-3}$ of Sec.~\ref{sec:caps} was earned by
\emph{clustering}: every within-cap pair sits near coincidence, where the astrometric
correlation is at its maximum. A constellation tiling the whole sky gives up exactly that.
Its pairs are isotropically distributed, so the pair average runs over the whole curve
$\Gamma(\Theta)$ of Sec.~\ref{sec:caps}, weighted by the separation density
$\tfrac12\sin\Theta$:
\begin{equation}
  \overline{\Gamma^2}_{\rm sky}
  =\int_0^\pi\!2\,\Gamma(\Theta)^2\,\tfrac12\sin\Theta\,d\Theta
  \simeq2.0\times10^{-3},
  \label{eq:gbarsky}
\end{equation}
a factor $3.5$ below the clustered-cap value, for both full-sky configurations. The
loss is milder than the zero crossing near $57^\circ$ might suggest, because the
anticorrelated lobe beyond it covers most of the solid angle and supplies $57\%$ of
$\overline{\Gamma^2}_{\rm sky}$: anticorrelation is as useful to a cross-correlation as
correlation. Since $\Omega_0\propto1/\sqrt{\overline{\Gamma^2}}$, it costs a
factor $\simeq1.9$, so the net gain of going full-sky is not the naive $250$ in $N$ but
$250/1.9\simeq130$.

\subsection{A next-generation aperture}
\label{sec:aperture}
The noise floor is set by photon-limited centroiding,
$\sigma_c\simeq\sigma_{\rm PSF}/\sqrt{N_{\rm ph}}\propto\lambda/D^2$ --- a larger aperture both
sharpens the diffraction PSF ($\sigma_{\rm PSF}=\lambda/D$) and collects more photons
($N_{\rm ph}\propto D^2$) --- so
\begin{equation}
  \Sigma \;\propto\; \frac{\lambda^2}{D^4},
  \qquad
  \Omega_0^{\rm thr}\;\propto\;\frac{\Sigma}{N}\;\propto\;\frac{\lambda^2}{N\,D^4},
  \label{eq:scaling}
\end{equation}
and the threshold improves as the fourth power of the aperture, which is why aperture, not
depth, is the lever. Our next-generation concept is an $8$-m aperture observing in the
near-UV ($\lambda\simeq0.3\,\mu$m), in the \emph{same} polar-cap geometry and at the
\emph{same} $G=13$, so with the same $N=10^4$ references and the same $\overline{\Gamma^2}$:
it lowers $\Sigma$ by $(\lambda/\lambda_{\rm Gaia})^2(D_{\rm Gaia}/D)^4$, more than three
orders of magnitude.

\section{Numerical estimates}
\label{sec:numbers}

\begin{table}[!t]
\caption{As Table~\ref{tab:thresholds} but for the next-generation aperture ($8$-m,
$\lambda\!\simeq\!0.3\,\mu$m, hence $\sigma_c=0.52\,\mu$as and
$\Sigma\simeq2.7\times10^{-4}\,\Sigma_{\rm Gaia}=2.8\times10^{-19}$~s), in the
\emph{same} polar-cap geometry and at the \emph{same} $G=13$, so with the same $N=10^4$ and
$\overline{\Gamma^2}=6.8\times10^{-3}$. The two tables therefore differ by the aperture alone.
Threshold $\Omega_0$ at $\mathrm{SNR}=1$, $f_{\rm Ny}=1$~mHz.}
\begin{ruledtabular}
\begin{tabular}{lc}
 $\alpha$ & $\Omega_0$ ($N=10^4$, $8$-m) \\
\colrule
 $-1$ \ (scale-inv., $\gamma=5$)   & $3.0\times10^{-11}$ \\
 $-2/3$ \ (SMBHB, $\gamma=13/3$)   & $1.2\times10^{-10}$ \\
 $-1/3$ \ ($\gamma=11/3$)          & $4.2\times10^{-10}$ \\
 $0$ \ ($\gamma=3$)                & $1.2\times10^{-9}$ \\
 $1$ \ (blue, $\gamma=1$)          & $1.1\times10^{-15}$ \\
\end{tabular}
\end{ruledtabular}
\label{tab:nextgen}
\end{table}

With the noise floor of Eq.~\eqref{eq:SigmaG} and the configurations of
Sec.~\ref{sec:configs} fixed, we now put numbers on the reach.

The numerical results come from a Python notebook (NumPy, SciPy, Matplotlib)
whose code was written by the AI model Claude Opus~5.5 (Anthropic, model ID
\texttt{claude-opus-5-5}), used through the Claude Code agentic coding interface. The authors
specified the quantities to compute, the equations of Secs.~\ref{sec:noise}--\ref{sec:configs}
and the parameters, and reviewed the code. At their direction the notebook also checks the
numerics against the analytic results, and recomputes every number printed in the paper and compares it with
the printed value.

\subsection{Thresholds and reach}
\label{sec:reach}
Each entry of Tables~\ref{tab:thresholds} and~\ref{tab:nextgen} is the amplitude at the pivot
for which the full-band integral of Eq.~\eqref{eq:snr_integral} reaches $\mathrm{SNR}=1$; the
index $\alpha$ decides where in the band the detection is made, and entries rescale between
configurations as $\Omega_0\propto\Sigma/(\sqrt{\overline{\Gamma^2}}\,N)$
[Eq.~\eqref{eq:scaling}]. The pivot is a poor place to describe blue spectra: an
$\Omega\propto f^4$ spectrum grows by $(f_{\rm Ny}/f_{\rm ref})^4\sim10^{18}$ between $30$~nHz
and $1$~mHz, which is why the $\alpha=1$ rows look so small. The same pivot effect makes
$\Omega_0$ non-monotonic in $\alpha$: red spectra ($\alpha<1/4$) are detected at
$1/T\simeq3$~nHz, below $f_{\rm ref}$, where a bluer spectrum carries less power for the same
$\Omega_0$, while past $\alpha=1/4$ the detection jumps to $f_{\rm Ny}$; the worst case, at
$\alpha\approx0.14$ with $\Omega_0=5.6\times10^{-6}$, is a choice of pivot, not a physical
loss of sensitivity.

The same divide sets where the spin pays off. In the noise-dominated regime the SNR
integrand is $f^{4\alpha-2}$, so $\int_{1/T}^{f_{\rm Ny}}$ is dominated by its lower limit for
$\alpha<1/4$ and by its upper limit for $\alpha>1/4$. For red backgrounds the information
sits in the lowest bins and spinning faster changes nothing; for blue or flat-in-band ones
every doubling of the spin rate appends a fresh, dominant octave below $f_{\rm Ny}$, hence
the power-law gain of Eq.~\eqref{eq:blue}. The spin cannot be raised without limit: once
the per-transit time $t_{\rm int}=\eta/f_s$ [Eq.~\eqref{eq:tint}] falls to the $\sim$second
of a Gaia-like focal-plane crossing, centroiding leaves the photon-limited regime and
$\Sigma$ is no longer spin-independent. This caps $f_s\lesssim2$~mHz (period $\sim\!8$~min,
$\sim\!40\times$ faster than Gaia), i.e.\ $f_{\rm Ny}\simeq1$~mHz, which we adopt as the
upper edge of the band. Raising the spin from Gaia-like ($f_{\rm Ny}\simeq2.3\times10^{-5}$~Hz)
to this ceiling, at $N=10^4$ and $G=13$, gives
\begin{align*}
 \alpha=\tfrac12:&\quad \Omega_0\!:\,5.0\times10^{-7}\!\to\!7.6\times10^{-8}\ (\times6.6),\\
 \alpha=1:&\quad \Omega_0\!:\,1.1\times10^{-9}\!\to\!4.0\times10^{-12}\ (\times2.8\!\times\!10^{2}),
\end{align*}
and a continuous source at $f_{\rm GW}$ is detectable only once $f_{\rm Ny}>f_{\rm GW}$:
cadence is a hard threshold, not a scaling.

The PTA amplitude $A\simeq2.4\times10^{-15}$ at
$f_{\rm yr}$~\cite{NANOGrav15yr,EPTADR2,PPTADR3,CPTADR1} corresponds to
$\Omega_{\rm GW}(30\,\text{nHz})\simeq7.1\times10^{-9}$
[$S_h\simeq2.1\times10^{-22}\,\text{Hz}^{-1}$; $7.4\times10^{-9}$ at $f_{\rm yr}$], and the
four configurations of Fig.~\ref{fig:sens} compare with it as follows. The Gaia-class polar
caps sit a factor $61$ above it on the SMBHB background and, since $\Omega_0\propto1/N$,
would meet it only at $N\simeq6\times10^5$, an order of magnitude beyond the
$\sim\!4\times10^4$ stars in the caps. The $8$-m aperture lowers every threshold by more than
three orders of magnitude (Table~\ref{tab:nextgen}, orange), taking the survey below current
PTAs (EPTA) at the bottom of the band, though not to the depth of an SKA-era array. The
Gaia-class constellation ($N=2.5\times10^{6}$,
$\overline{\Gamma^2}_{\rm sky}=2.0\times10^{-3}$) reaches
\begin{equation}
  \Omega_0^{\rm SMBHB}\!\simeq\!3.2\times10^{-9},\qquad
  \Omega_0^{\text{scale-inv}}\!\simeq\!8.4\times10^{-10},
  \label{eq:constthr}
\end{equation}
only about $2.2\times$ below the PTA amplitude (green dashed): once the loss of overlap is
charged for, multiplicity alone does not beat pulsar timing at the bottom of the band, and
aperture is the lever that does. Combining both, a constellation of $8$-m instruments
reaches
\begin{equation}
  \Omega_0^{\rm SMBHB}\!\simeq\!8.7\times10^{-13},\qquad
  \Omega_0^{\text{scale-inv}}\!\simeq\!2.3\times10^{-13},
  \label{eq:ultimate}
\end{equation}
some $130\times$ below a single $8$-m aperture and more than three orders of magnitude below
the Gaia-class constellation --- $\simeq\!5.0\times10^{5}$ over the fiducial survey --- and
approaches an SKA-era array at the bottom of the band while extending continuously to
$1$~mHz (purple dot-dashed): a wide-band, all-sky $\mu$Hz observatory, if an indicative one,
some twenty satellites spread over half the Earth's orbit being a major investment. In every
configuration the strain reach stays well above that of a dedicated $\mu$Hz interferometer;
the value of the concept lies in covering the $\mu$Hz gap in frequency and in the flat and
blue regime opened by fast cadence, not in duplicating PTAs at the bottom of the band.

Panel (b) of Fig.~\ref{fig:sens} is the companion view: the PLS of Eq.~\eqref{eq:pls}, built
from the same $S_{\rm eff}$ and reproducing Table~\ref{tab:thresholds} at $\beta=2\alpha+2$ to
better than $1\%$. Its astrometric curves are nearly flat, the signature of a white
deflection floor integrated over a wide band, whereas the PTA curves rise as $f^{3/2}$: the
Gaia-class polar-cap curve drops below EPTA at $10$~nHz, a factor $3$ above the band edge
(forecasts for absolute Gaia astrometry place this crossing above
$\sim\!10^{-7.5}$~Hz~\cite{Zhang2025,Klioner2018}), and beyond the PTA Nyquist limit nothing
else reaches the band. For red spectra the SNR still accrues at the bottom of the band, where
the lever is the baseline rather than the cadence; in $h_c$ the minimum of panel (b) sits at
$f\simeq8$~nHz.

\subsection{Systematics and low-frequency floor}
The global astrometric fit acts as a high-pass filter: offsets, drifts and the annual
parallax are absorbed star by star, so bins below $1/T$ carry no GW information and we
integrate from $1/T$~\cite{Geyer2025}. This makes the survey complementary to the Gaia DR3
quasar analysis~\cite{Akhmetov2026}, which constrains the secular drift, i.e.\ frequencies
below $1/2T_{\rm DR3}\simeq5.6$~nHz (minimum detectable strain $\simeq10^{-11}$, and
$\simeq3\times10^{-12}$ forecast for DR4), and to its joint analysis with binary-pulsar
timing~\cite{Filipello2026}. Real astrometry is also not purely photon-limited: attitude and
calibration errors are spatially correlated and are separated in the
vector-spherical-harmonic basis, where the $\ell=1$ modes --- dominated by the secular
aberration drift and residual frame spin --- are marginalized while the GW background lives
in $\ell\ge2$~\cite{Akhmetov2026,Mihaylov2018}. Together with stellar jitter (activity,
granulation, unresolved companions), this makes the photon-limited $\Sigma$ an optimistic
floor.

The polar-cap geometry adds a mission-dependent caveat. An attitude solution referenced to
the science stars absorbs the deflection common to the field --- the ``mean-signal
subtraction'' that dominates the Roman and \emph{Kepler} budgets~\cite{Zhang2025} --- leaving
only the differential response within the field~\cite{Geyer2026}. This bites where our
signal lives, in near-coincident within-cap pairs. Two features limit the damage. A single
attitude has only two tilt degrees of freedom for the four components of the two caps' mean
displacements, which are independent since $T(\pi)=0$, so a common tilt removes only about
half of each cap's coherent signal, while the within-cap divergence and shear survive
regardless; and an attitude reference independent of the science field (drag-free
metrology, a basic-angle monitor, or a global solution with no power in the band) loses
little. We have not folded this into the numbers: it is the assumption most likely to move
$\overline{\Gamma^2}$ in either direction.

\subsection{Putting the PTAs on the same footing}
Every curve in Fig.~\ref{fig:sens} uses the same estimator, Eq.~\eqref{eq:snr_opt} at
$\mathrm{SNR}=1$ in one bin $\Delta f=1/T_{\rm obs}$ in the noise-dominated limit,
\begin{equation}
  h_c^{\rm sens}(f)=\big[f\,S_{\rm eff}(f)\big]^{1/2},
  \qquad
  S_{\rm eff}=\frac{S_n}{\overline{\Gamma}\,N_{\rm src}},
  \label{eq:hcsens}
\end{equation}
in the notation of Eq.~\eqref{eq:seff}: $N_{\rm src}$ sources of strain-referred noise $S_n$,
with $S_n=\Sigma$ and $N_{\rm src}=N$ for the survey. For a PTA the white timing-residual PSD
$2\sigma_p^2\Delta_p$, of the same form as Eq.~\eqref{eq:Sigma}, is referred to strain as
$S_{n,P}=12\pi^2f^2(2\sigma_p^2\Delta_p)$ [white part of Eq.~(20) of
Ref.~\cite{Vaglio2025}] --- the $f^2$ being the squared $1/f$ of the timing observable ---
and the response is the Hellings--Downs function
$\mu(\zeta)=\tfrac12-\tfrac{x}{4}+\tfrac32x\ln x$, $x=(1-\cos\zeta)/2$, whose isotropic pair
average gives
\begin{equation}
  \overline{\Gamma^2}_{\rm PTA}=\langle\mu^2\rangle=\tfrac1{48}\simeq2.08\times10^{-2},
  \label{eq:hdbar}
\end{equation}
within a factor $1.75$ in amplitude of the polar caps' $\overline{\Gamma}$; unit overlap
would flatter the PTA curves by a further $2.6$.

The PTA curves are deliberately toy models: single white-noise arrays with EPTA ($M=25$,
$\sigma_p=11\,\mu$s, $\Delta_p=33$~d) and SKA ($M=2500$, $\sigma_p=100$~ns, $\Delta_p=14$~d)
parameters from Ref.~\cite{Vaglio2025}, each with $h_c^{\rm sens}\propto f^{3/2}$ over
$[1/T_{\rm obs},1/2\Delta_p]$; the real EPTA analysis uses decades of heterogeneous data and
red-noise modelling. EPTA is drawn on its real $24.7$-yr DR2 baseline: on $10$~yr the same
array reaches only $\Omega_\beta=1.1\times10^{-7}$ at $\beta=2/3$, a factor $16$ above the
measured amplitude, while on $24.7$~yr it reaches $2.3\times10^{-9}$, recovering the
background at $\mathrm{SNR}\simeq3$ as EPTA reports~\cite{EPTADR2}. The gain is steep because
a PTA's strain noise falls as $f^2$ toward low frequency, so
$\mathrm{SNR}\propto T^{3-2\alpha}=T^{13/3}$, a factor $\simeq50$. It shows only in panel
(b): the per-bin $h_c^{\rm sens}$ is $T$-independent, so a longer baseline merely extends
panel (a) to lower $f$. LISA is the analytic single-detector curve of Ref.~\cite{Robson2019}.

\section{Conclusions}
\label{sec:conclusions}

We have forecast the SGWB sensitivity of a dedicated, fast-spinning astrometric
survey targeting the microhertz gap. Because the astrometric deflection measures the
strain $h$ directly, the per-source noise is white in frequency and the single-epoch
error grows as $\sqrt{f_s}$ in exactly the way that keeps the noise floor $\Sigma$
independent of spin: a faster spin buys observable bandwidth --- up to a
photon-limited Nyquist ceiling $f_{\rm Ny}\!\simeq\!1$~mHz --- rather than noise. Over a
common $10$-yr baseline a Gaia-class instrument ($G\!\simeq\!13$, $N\!\sim\!10^4$,
$\overline{\Gamma^2}\!\simeq\!6.8\times10^{-3}$) sits a factor
$\sim\!60$ above the current PTA
amplitude on a red SMBHB background, which it would match only at
$N\!\simeq\!6\times10^5$ references --- beyond a single pair of caps, and a reminder
that the concept's value is not in duplicating PTAs at the bottom of the band. For flat
or blue spectra the fast cadence instead converts directly into reach across a band no
interferometer yet covers. Because $\Omega_0\propto\Sigma/N\propto10^{0.05G}$, depth is
neutral and the concept has exactly two levers: aperture and sky coverage. Holding the
magnitude fixed at $G=13$, an $8$-m aperture in the same polar caps gains more than three
orders of magnitude in $\Omega_0$ and would beat present PTAs (EPTA) at the nanohertz end,
while an all-sky constellation of Gaia-class satellites
($N=2.5\times10^{6}$) gains only $\simeq\!130$ once the loss of the clustered-cap overlap is
charged for, and so passes only modestly the current PTA
amplitude. Combining the two gives
$\Omega_0^{\rm SMBHB}\simeq8.7\times10^{-13}$, approaching an SKA-era array compared on the
same footing, and extends continuously across the whole nHz--mHz
range as a single wide-band survey, while flagging the loudest resolved SMBHB coalescences
as multimessenger targets. That the aperture lever beats the multiplicity lever by more
than an order of magnitude is, we think, the most useful design conclusion here.

Beyond the background, the $\mu$Hz band is where binaries of $\sim\!10^7$--$10^8\,M_\odot$
coalesce [$f_{\rm ISCO}\simeq0.4\,\text{mHz}\,(10^7M_\odot/M)$]: the population PTAs see only
as a background~\cite{SesanaSMBHB2008,Criswell2026}, and a central target of the proposed
$\mu$Hz probes~\cite{SesanaMicroHz2021,Fedderke2022,JenkinsBlasFoster2026}. Astrometry
records the full two-dimensional deflection, and hence the polarization of a resolved
binary~\cite{BookFlanagan2011,Mihaylov2018,Moore2017}, although localizing it needs the sky
coverage of the constellation rather than two caps~\cite{Zhang2025}. Only the loudest such
events would be resolved, and searching for them needs precomputed data--template inner
products over a terabyte-scale data set~\cite{Zhang2025} and frequency-evolving templates
that are only now being developed~\cite{ZhangPardoBecsy2026}.

These estimates are deliberately optimistic in their noise model: the photon-limited
floor neglects the residual attitude and calibration systematics and the stellar
jitter that a flown mission must control. They are also optimistic in crediting both
deflection components with the precision $\sigma_c$: a scanning transit measures the
along-scan coordinate precisely and the across-scan one only coarsely. Because the scan
orientation across each cap turns only once a year, nearby stars are measured along nearly
the same direction at any epoch, so a pair contributes one of the two equal terms of
$\sum_{a,b}(\Gamma^{ab})^2=2\,\Gamma(\Theta)^2$ rather than both. In the worst case, an
across-scan measurement with no astrometric value, this halves $\overline{\Gamma^2}$ and
raises every threshold by at most $\sqrt2\simeq1.4$ in $\Omega_0$ and $2^{1/4}\simeq1.2$ in
$h_c^{\rm sens}$ --- well within the other uncertainties of the forecast, and partly
recoverable with a focal plane that centroids in two dimensions. Both astrometric overlaps are computed rather than assumed, from the closed-form
pair correlation of Ref.~\cite{Mihaylov2018}, which puts them on the same exact footing as
the PTA value $1/48$: $\overline{\Gamma^2}\simeq6.8\times10^{-3}$ for the two caps, half the
coincidence ceiling because antipodal stars are uncorrelated, and
$\overline{\Gamma^2}_{\rm sky}\simeq2.0\times10^{-3}$ for the full sky. What remains
uncertain is how much of the near-coincident correlation survives the attitude solution
(Sec.~\ref{sec:numbers}), which depends on the mission design. None of this overturns the
central point --- direct astrometry is the natural high-frequency probe of the $\mu$Hz gap, and
cadence is the lever that opens
it.

\begin{acknowledgments}
GM acknowledges support from the
ANR grant AAPG2024 PRC - GalaxyFIT. Claude Opus~5.5 (Anthropic) was used to help with the writing: it drafted and revised
passages from the authors' notes and instructions, proofread the text for typos and
inconsistencies, and searched the web for additional literature, all of which the authors
checked against the original papers before citing. The authors take full
responsibility for the content of the paper.
\end{acknowledgments}

\bibliography{paper}

@article{NANOGrav15yr,
  author  = {{Agazie}, Gabriella and others},
  collaboration = {NANOGrav},
  title   = {{The NANOGrav 15 yr Data Set: Evidence for a Gravitational-wave Background}},
  journal = {Astrophys. J. Lett.},
  volume  = {951},
  number  = {1},
  pages   = {L8},
  year    = {2023},
  doi     = {10.3847/2041-8213/acdac6},
  eprint  = {2306.16213},
  archivePrefix = {arXiv},
  primaryClass  = {astro-ph.HE},
}

@article{EPTADR2,
  author  = {{Antoniadis}, J. and others},
  collaboration = {EPTA and InPTA},
  title   = {{The second data release from the European Pulsar Timing Array. III. Search for gravitational wave signals}},
  journal = {Astron. Astrophys.},
  volume  = {678},
  pages   = {A50},
  year    = {2023},
  doi     = {10.1051/0004-6361/202346844},
  eprint  = {2306.16214},
  archivePrefix = {arXiv},
  primaryClass  = {astro-ph.HE},
}

@article{PPTADR3,
  author  = {{Reardon}, Daniel J. and others},
  collaboration = {PPTA},
  title   = {{Search for an Isotropic Gravitational-wave Background with the Parkes Pulsar Timing Array}},
  journal = {Astrophys. J. Lett.},
  volume  = {951},
  number  = {1},
  pages   = {L6},
  year    = {2023},
  doi     = {10.3847/2041-8213/acdd02},
  eprint  = {2306.16215},
  archivePrefix = {arXiv},
  primaryClass  = {astro-ph.HE},
}

@article{CPTADR1,
  author  = {{Xu}, Heng and others},
  collaboration = {CPTA},
  title   = {{Searching for the Nano-Hertz Stochastic Gravitational Wave Background with the Chinese Pulsar Timing Array Data Release I}},
  journal = {Res. Astron. Astrophys.},
  volume  = {23},
  number  = {7},
  pages   = {075024},
  year    = {2023},
  doi     = {10.1088/1674-4527/acdfa5},
  eprint  = {2306.16216},
  archivePrefix = {arXiv},
  primaryClass  = {astro-ph.HE},
}

@article{Vaglio2025,
  author  = {{Vaglio}, Massimo and {Falxa}, Mikel and {Mentasti}, Giorgio and {Renzini}, Arianna I. and {Kuntz}, Adrien and {Barausse}, Enrico and {Contaldi}, Carlo R. and {Sesana}, Alberto},
  title   = {{Searching for gravitational waves with Gaia and its cross-correlation with PTA: Absolute vs differential astrometry}},
  journal = {Phys. Rev. D},
  volume  = {113},
  number  = {12},
  pages   = {124003},
  year    = {2026},
  doi     = {10.1103/cg1z-2r29},
  eprint  = {2507.18593},
  archivePrefix = {arXiv},
  primaryClass  = {gr-qc},
}

@article{MentastiContaldi2024,
  author  = {{Mentasti}, Giorgio and {Contaldi}, Carlo R.},
  title   = {{Observing gravitational waves with solar system astrometry}},
  journal = {JCAP},
  volume  = {2024},
  number  = {05},
  pages   = {028},
  year    = {2024},
  doi     = {10.1088/1475-7516/2024/05/028},
  eprint  = {2311.03474},
  archivePrefix = {arXiv},
  primaryClass  = {gr-qc},
}

@article{Mihaylov2018,
  author  = {{Mihaylov}, Deyan P. and {Moore}, Christopher J. and {Gair}, Jonathan R. and {Lasenby}, Anthony and {Gilmore}, Gerard},
  title   = {{Astrometric effects of gravitational wave backgrounds with non-Einsteinian polarizations}},
  journal = {Phys. Rev. D},
  volume  = {97},
  number  = {12},
  pages   = {124058},
  year    = {2018},
  doi     = {10.1103/PhysRevD.97.124058},
  eprint  = {1804.00660},
  archivePrefix = {arXiv},
  primaryClass  = {gr-qc},
}

@article{Inomata2024,
  author  = {{Inomata}, Keisuke and {Kamionkowski}, Marc and {Toral}, Celia M. and {Taylor}, Stephen R.},
  title   = {{Overlap reduction functions for pulsar timing arrays and astrometry}},
  journal = {Phys. Rev. D},
  volume  = {110},
  number  = {6},
  pages   = {063547},
  year    = {2024},
  doi     = {10.1103/PhysRevD.110.063547},
  eprint  = {2406.00096},
  archivePrefix = {arXiv},
  primaryClass  = {astro-ph.CO},
}

@article{Caliskan2024,
  author  = {{\c{C}al{\i}\c{s}kan}, Mesut and {Chen}, Yifan and {Dai}, Liang and {Anil Kumar}, Neha and {Stomberg}, Isak and {Xue}, Xiao},
  title   = {{Dissecting the stochastic gravitational wave background with astrometry}},
  journal = {JCAP},
  volume  = {2024},
  number  = {05},
  pages   = {030},
  year    = {2024},
  doi     = {10.1088/1475-7516/2024/05/030},
  eprint  = {2312.03069},
  archivePrefix = {arXiv},
  primaryClass  = {astro-ph.CO},
}

@incollection{Lindegren1978,
  author    = {{Lindegren}, Lennart},
  title     = {{Photoelectric astrometry -- A comparison of methods for precise image location}},
  booktitle = {Modern Astrometry, IAU Colloq. 48},
  editor    = {{Prochazka}, F. V. and {Tucker}, R. H.},
  pages     = {197},
  year      = {1978},
  publisher = {University Observatory Vienna},
}

@article{GaiaMission2016,
  author  = {{Gaia Collaboration} and {Prusti}, T. and others},
  title   = {{The Gaia mission}},
  journal = {Astron. Astrophys.},
  volume  = {595},
  pages   = {A1},
  year    = {2016},
  doi     = {10.1051/0004-6361/201629272},
  eprint  = {1609.04153},
  archivePrefix = {arXiv},
  primaryClass  = {astro-ph.IM},
}

@article{Fabricius2016,
  author  = {{Fabricius}, C. and {Bastian}, U. and {Portell}, J. and others},
  title   = {{Gaia Data Release 1. Pre-processing and source list creation}},
  journal = {Astron. Astrophys.},
  volume  = {595},
  pages   = {A3},
  year    = {2016},
  doi     = {10.1051/0004-6361/201628643},
  eprint  = {1609.04273},
  archivePrefix = {arXiv},
  primaryClass  = {astro-ph.IM},
}

@article{LVKGWTC3,
  author  = {{Abbott}, R. and others},
  collaboration = {LIGO Scientific, Virgo and KAGRA},
  title   = {{GWTC-3: Compact Binary Coalescences Observed by LIGO and Virgo during the Second Part of the Third Observing Run}},
  journal = {Phys. Rev. X},
  volume  = {13},
  number  = {4},
  pages   = {041039},
  year    = {2023},
  doi     = {10.1103/PhysRevX.13.041039},
  eprint  = {2111.03606},
  archivePrefix = {arXiv},
  primaryClass  = {gr-qc},
}

@article{SesanaSMBHB2008,
  author  = {{Sesana}, A. and {Vecchio}, A. and {Colacino}, C. N.},
  title   = {{The stochastic gravitational-wave background from massive black hole binary systems: implications for observations with Pulsar Timing Arrays}},
  journal = {Mon. Not. R. Astron. Soc.},
  volume  = {390},
  number  = {1},
  pages   = {192},
  year    = {2008},
  doi     = {10.1111/j.1365-2966.2008.13682.x},
  eprint  = {0804.4476},
  archivePrefix = {arXiv},
  primaryClass  = {astro-ph},
}

@article{CapriniFigueroa2018,
  author  = {{Caprini}, Chiara and {Figueroa}, Daniel G.},
  title   = {{Cosmological backgrounds of gravitational waves}},
  journal = {Class. Quantum Grav.},
  volume  = {35},
  number  = {16},
  pages   = {163001},
  year    = {2018},
  doi     = {10.1088/1361-6382/aac608},
  eprint  = {1801.04268},
  archivePrefix = {arXiv},
  primaryClass  = {astro-ph.CO},
}

@article{Christensen2019,
  author  = {{Christensen}, Nelson},
  title   = {{Stochastic gravitational wave backgrounds}},
  journal = {Rep. Prog. Phys.},
  volume  = {82},
  number  = {1},
  pages   = {016903},
  year    = {2019},
  doi     = {10.1088/1361-6633/aae6b5},
  eprint  = {1811.08797},
  archivePrefix = {arXiv},
  primaryClass  = {gr-qc},
}

@article{LISA2017,
  author  = {{Amaro-Seoane}, Pau and others},
  title   = {{Laser Interferometer Space Antenna}},
  journal = {arXiv e-prints},
  year    = {2017},
  eprint  = {1702.00786},
  archivePrefix = {arXiv},
  primaryClass  = {astro-ph.IM},
}

@article{SesanaMicroHz2021,
  author  = {{Sesana}, Alberto and others},
  title   = {{Unveiling the gravitational universe at $\mu$-Hz frequencies}},
  journal = {Exp. Astron.},
  volume  = {51},
  number  = {3},
  pages   = {1333},
  year    = {2021},
  doi     = {10.1007/s10686-021-09709-9},
  eprint  = {1908.11391},
  archivePrefix = {arXiv},
  primaryClass  = {astro-ph.CO},
}

@article{HellingsDowns1983,
  author  = {{Hellings}, R. W. and {Downs}, G. S.},
  title   = {{Upper limits on the isotropic gravitational radiation background from pulsar timing analysis}},
  journal = {Astrophys. J. Lett.},
  volume  = {265},
  pages   = {L39},
  year    = {1983},
  doi     = {10.1086/183954},
}

@article{Detweiler1979,
  author  = {{Detweiler}, S.},
  title   = {{Pulsar timing measurements and the search for gravitational waves}},
  journal = {Astrophys. J.},
  volume  = {234},
  pages   = {1100},
  year    = {1979},
  doi     = {10.1086/157593},
}

@article{Pyne1996,
  author  = {{Pyne}, Ted and {Gwinn}, Carl R. and {Birkinshaw}, Mark and {Eubanks}, T. Marshall and {Matsakis}, Demetrios N.},
  title   = {{Gravitational Radiation and Very Long Baseline Interferometry}},
  journal = {Astrophys. J.},
  volume  = {465},
  pages   = {566},
  year    = {1996},
  doi     = {10.1086/177443},
}

@article{BookFlanagan2011,
  author  = {{Book}, Laura G. and {Flanagan}, \'Eanna \'E.},
  title   = {{Astrometric Effects of a Stochastic Gravitational Wave Background}},
  journal = {Phys. Rev. D},
  volume  = {83},
  number  = {2},
  pages   = {024024},
  year    = {2011},
  doi     = {10.1103/PhysRevD.83.024024},
  eprint  = {1009.4192},
  archivePrefix = {arXiv},
  primaryClass  = {astro-ph.CO},
}

@article{Klioner2018,
  author  = {{Klioner}, Sergei A.},
  title   = {{Gaia-like astrometry and gravitational waves}},
  journal = {Class. Quantum Grav.},
  volume  = {35},
  number  = {4},
  pages   = {045005},
  year    = {2018},
  doi     = {10.1088/1361-6382/aa9f57},
  eprint  = {1710.11474},
  archivePrefix = {arXiv},
  primaryClass  = {astro-ph.HE},
}

@article{Theia2017,
  author  = {{The Theia Collaboration} and {Boehm}, Celine and others},
  title   = {{Theia: Faint objects in motion or the new astrometry frontier}},
  journal = {arXiv e-prints},
  year    = {2017},
  eprint  = {1707.01348},
  archivePrefix = {arXiv},
  primaryClass  = {astro-ph.IM},
}

@article{Darling2018,
  author  = {{Darling}, Jeremy and {Truebenbach}, Alexandra E. and {Paine}, Jennie},
  title   = {{Astrometric Limits on the Stochastic Gravitational Wave Background}},
  journal = {Astrophys. J.},
  volume  = {861},
  number  = {2},
  pages   = {113},
  year    = {2018},
  doi     = {10.3847/1538-4357/aac772},
  eprint  = {1804.06986},
  archivePrefix = {arXiv},
  primaryClass  = {astro-ph.CO},
}

@article{PardoRoman2023,
  author  = {{Pardo}, Kris and {Chang}, Tzu-Ching and {Dor\'e}, Olivier and {Wang}, Yijun},
  title   = {{Gravitational Wave Detection with Relative Astrometry using Roman's Galactic Bulge Time Domain Survey}},
  journal = {arXiv e-prints},
  year    = {2023},
  eprint  = {2306.14968},
  archivePrefix = {arXiv},
  primaryClass  = {astro-ph.CO},
}

@article{Akhmetov2026,
  author  = {{Akhmetov}, V. and {Filipello}, L. and {Crosta}, M. and {Lattanzi}, M. G. and {Bucciarelli}, B. and {Abbas}, U. and {Santucci}, F.},
  title   = {{Low-Frequency Stochastic Gravitational-Wave Background in Gaia DR3 catalogue}},
  journal = {arXiv e-prints},
  year    = {2026},
  eprint  = {2603.23028},
  archivePrefix = {arXiv},
  primaryClass  = {astro-ph.CO},
}

@article{Fedderke2022,
  author  = {{Fedderke}, Michael A. and {Graham}, Peter W. and {Rajendran}, Surjeet},
  title   = {{Asteroids for $\mu$Hz gravitational-wave detection}},
  journal = {Phys. Rev. D},
  volume  = {105},
  number  = {10},
  pages   = {103018},
  year    = {2022},
  doi     = {10.1103/PhysRevD.105.103018},
  eprint  = {2112.11431},
  archivePrefix = {arXiv},
  primaryClass  = {gr-qc},
}

@article{BlasJenkins2022,
  author  = {{Blas}, Diego and {Jenkins}, Alexander C.},
  title   = {{Bridging the $\mu$Hz Gap in the Gravitational-Wave Landscape with Binary Resonances}},
  journal = {Phys. Rev. Lett.},
  volume  = {128},
  number  = {10},
  pages   = {101103},
  year    = {2022},
  doi     = {10.1103/PhysRevLett.128.101103},
  eprint  = {2107.04601},
  archivePrefix = {arXiv},
  primaryClass  = {gr-qc},
}

@article{Qin2019,
  author  = {{Qin}, Wenzer and {Boddy}, Kimberly K. and {Kamionkowski}, Marc and {Dai}, Liang},
  title   = {{Pulsar-timing arrays, astrometry, and gravitational waves}},
  journal = {Phys. Rev. D},
  volume  = {99},
  number  = {6},
  pages   = {063002},
  year    = {2019},
  doi     = {10.1103/PhysRevD.99.063002},
  eprint  = {1810.02369},
  archivePrefix = {arXiv},
  primaryClass  = {astro-ph.CO},
}

@article{Jaraba2023,
  author  = {{Jaraba}, Santiago and {Garc\'ia-Bellido}, Juan and {Kuroyanagi}, Sachiko and {Ferraiuolo}, Sarah and {Braglia}, Matteo},
  title   = {{Stochastic gravitational wave background constraints from Gaia DR3 astrometry}},
  journal = {Mon. Not. R. Astron. Soc.},
  volume  = {524},
  number  = {3},
  pages   = {3609},
  year    = {2023},
  doi     = {10.1093/mnras/stad2141},
  eprint  = {2304.06350},
  archivePrefix = {arXiv},
  primaryClass  = {astro-ph.CO},
}

@article{Robson2019,
  author  = {{Robson}, Travis and {Cornish}, Neil J. and {Liu}, Chang},
  title   = {{The construction and use of LISA sensitivity curves}},
  journal = {Class. Quantum Grav.},
  volume  = {36},
  number  = {10},
  pages   = {105011},
  year    = {2019},
  doi     = {10.1088/1361-6382/ab1101},
  eprint  = {1803.01944},
  archivePrefix = {arXiv},
  primaryClass  = {astro-ph.IM},
}

@article{ThraneRomano2013,
  author  = {{Thrane}, Eric and {Romano}, Joseph D.},
  title   = {{Sensitivity curves for searches for gravitational-wave backgrounds}},
  journal = {Phys. Rev. D},
  volume  = {88},
  number  = {12},
  pages   = {124032},
  year    = {2013},
  doi     = {10.1103/PhysRevD.88.124032},
  eprint  = {1310.5300},
  archivePrefix = {arXiv},
  primaryClass  = {astro-ph.IM},
}

@incollection{JenkinsBlasFoster2026,
  author    = {{Jenkins}, Alexander C. and {Blas}, Diego and {Foster}, Joshua W.},
  title     = {{Unveiling the Microhertz Gravitational-Wave Sky with the Square Kilometre Array Observatory}},
  booktitle = {{Advancing Astrophysics with the SKA II (AASKA II)}},
  year      = {2026},
  doi       = {10.48550/arXiv.2606.27418},
  eprint    = {2606.27418},
  archivePrefix = {arXiv},
  primaryClass  = {astro-ph.CO},
}

@article{Shimmering2024,
  author  = {{Mentasti}, Giorgio and {Contaldi}, Carlo R.},
  title   = {{Cosmic shimmering: the gravitational wave signal of time-resolved cosmic shear observations}},
  journal = {JCAP},
  volume  = {2025},
  number  = {06},
  pages   = {013},
  year    = {2025},
  doi     = {10.1088/1475-7516/2025/06/013},
  eprint  = {2410.16274},
  archivePrefix = {arXiv},
  primaryClass  = {astro-ph.CO},
}

@article{SpinHarmonic2025,
  author  = {{Mentasti}, Giorgio and {Contaldi}, Carlo R.},
  title   = {{A unified spin-harmonic framework for correlating pulsar timing, astrometric deflection, and shimmering gravitational wave observations}},
  journal = {JCAP},
  volume  = {2026},
  number  = {02},
  pages   = {013},
  year    = {2026},
  doi     = {10.1088/1475-7516/2026/02/013},
  eprint  = {2507.16906},
  archivePrefix = {arXiv},
  primaryClass  = {astro-ph.CO},
}

@article{FinkContaldiMentasti2025,
  author  = {{Fink}, Elias and {Contaldi}, Carlo R. and {Mentasti}, Giorgio},
  title   = {{Cross-correlating Astrometric and Timing Residuals to Constrain Stochastic Gravitational-Wave Backgrounds}},
  journal = {arXiv e-prints},
  year    = {2025},
  eprint  = {2510.25646},
  archivePrefix = {arXiv},
  primaryClass  = {gr-qc},
}

@article{Zhang2025,
  author  = {{Zhang}, Benjamin and {Pardo}, Kris and {Wang}, Yijun and {Bouma}, Luke and {Chang}, Tzu-Ching and {Dor\'e}, Olivier},
  title   = {{Fast Bayesian method for coherent gravitational wave searches with relative astrometry}},
  journal = {Phys. Rev. D},
  volume  = {112},
  number  = {4},
  pages   = {042002},
  year    = {2025},
  doi     = {10.1103/qzys-t232},
  eprint  = {2506.19206},
  archivePrefix = {arXiv},
  primaryClass  = {astro-ph.IM},
}

@article{Caldarola2026,
  author  = {{Caldarola}, Marienza and {Morr\'as}, Gonzalo and {Jaraba}, Santiago and {Kuroyanagi}, Sachiko and {Nesseris}, Savvas and {Garc\'ia-Bellido}, Juan},
  title   = {{Astrometric constraints on stochastic gravitational wave background with neural networks}},
  journal = {Phys. Rev. D},
  volume  = {113},
  number  = {4},
  pages   = {043522},
  year    = {2026},
  doi     = {10.1103/b6n8-gjwk},
  eprint  = {2412.15879},
  archivePrefix = {arXiv},
  primaryClass  = {astro-ph.CO},
}

@article{Wang2021,
  author  = {{Wang}, Yijun and {Pardo}, Kris and {Chang}, Tzu-Ching and {Dor\'e}, Olivier},
  title   = {{Gravitational Wave Detection with Photometric Surveys}},
  journal = {Phys. Rev. D},
  volume  = {103},
  number  = {8},
  pages   = {084007},
  year    = {2021},
  doi     = {10.1103/PhysRevD.103.084007},
  eprint  = {2010.02218},
  archivePrefix = {arXiv},
  primaryClass  = {gr-qc},
}

@article{Wang2022,
  author  = {{Wang}, Yijun and {Pardo}, Kris and {Chang}, Tzu-Ching and {Dor\'e}, Olivier},
  title   = {{Constraining the stochastic gravitational wave background with photometric surveys}},
  journal = {Phys. Rev. D},
  volume  = {106},
  number  = {8},
  pages   = {084006},
  year    = {2022},
  doi     = {10.1103/PhysRevD.106.084006},
  eprint  = {2205.07962},
  archivePrefix = {arXiv},
  primaryClass  = {gr-qc},
}

@article{ZhangPardoBecsy2026,
  author  = {{Zhang}, Benjamin and {Pardo}, Kris and {B\'ecsy}, Bence},
  title   = {{A compact time-domain reduced order quadrature for gravitational wave inspirals with non-stationary noise}},
  journal = {arXiv e-prints},
  year    = {2026},
  eprint  = {2609.14130},
  archivePrefix = {arXiv},
  primaryClass  = {gr-qc},
}

@article{Criswell2026,
  author  = {{Criswell}, Alexander W. and {Taylor}, Stephen R. and {Pardo}, Kris and {Sesana}, Alberto and {Izquierdo}, David and {Bonoli}, Silvia and {Spinoso}, Daniele},
  title   = {{Archival Multiband Gravitational-Wave Signals from Massive Black Hole Binary Mergers}},
  journal = {arXiv e-prints},
  year    = {2026},
  eprint  = {2604.21013},
  archivePrefix = {arXiv},
  primaryClass  = {astro-ph.HE},
}

@article{FedderkeStellar2022,
  author  = {{Fedderke}, Michael A. and {Graham}, Peter W. and {Macintosh}, Bruce and {Rajendran}, Surjeet},
  title   = {{Astrometric gravitational-wave detection via stellar interferometry}},
  journal = {Phys. Rev. D},
  volume  = {106},
  number  = {2},
  pages   = {023002},
  year    = {2022},
  doi     = {10.1103/PhysRevD.106.023002},
  eprint  = {2204.07677},
  archivePrefix = {arXiv},
  primaryClass  = {astro-ph.IM},
}

@article{Darling2025,
  author  = {{Darling}, Jeremy},
  title   = {{A New Approach to the Low-frequency Stochastic Gravitational-wave Background: Constraints from Quasars and the Astrometric Hellings--Downs Curve}},
  journal = {Astrophys. J. Lett.},
  volume  = {982},
  number  = {2},
  pages   = {L46},
  year    = {2025},
  doi     = {10.3847/2041-8213/adbf0d},
  eprint  = {2412.08605},
  archivePrefix = {arXiv},
  primaryClass  = {astro-ph.CO},
}

@article{Geyer2025,
  author  = {{Geyer}, R. and {Klioner}, S. A. and {Lindegren}, L. and {Lammers}, U.},
  title   = {{Influence of a continuous plane gravitational wave on Gaia-like astrometry}},
  journal = {Astron. Astrophys.},
  volume  = {695},
  pages   = {A172},
  year    = {2025},
  doi     = {10.1051/0004-6361/202453545},
  eprint  = {2412.15770},
  archivePrefix = {arXiv},
  primaryClass  = {astro-ph.IM},
}

@article{Geyer2026,
  author  = {{Geyer}, Robin and {Zschocke}, Sven and {Soffel}, Michael and {Klioner}, Sergei and {Lindegren}, Lennart and {Lammers}, Uwe},
  title   = {{Gravitational waves and small-field astrometry}},
  journal = {Phys. Rev. D},
  volume  = {113},
  number  = {4},
  pages   = {043021},
  year    = {2026},
  doi     = {10.1103/l59s-fz3z},
  eprint  = {2601.05754},
  archivePrefix = {arXiv},
  primaryClass  = {gr-qc},
}

@article{Perna2026,
  author  = {{Perna}, Gabriele and {Bellomo}, Nicola and {Roatti}, Vincenzo and {Bertacca}, Daniele},
  title   = {{From the solar system to cosmological distances: a complete formalism for gravitational wave astrometry}},
  journal = {JCAP},
  volume  = {2026},
  number  = {07},
  pages   = {102},
  year    = {2026},
  doi     = {10.1088/1475-7516/2026/07/102},
  eprint  = {2512.14668},
  archivePrefix = {arXiv},
  primaryClass  = {astro-ph.CO},
}

@article{PernaSKA2026,
  author  = {{Perna}, Gabriele and {Bellomo}, Nicola and {Bertacca}, Daniele},
  title   = {{Synergies Between Pulsar Timing Array and Astrometry}},
  journal = {arXiv e-prints},
  year    = {2026},
  eprint  = {2606.24992},
  archivePrefix = {arXiv},
  primaryClass  = {astro-ph.CO},
}

@article{JimenezCruz2025,
  author  = {{Jim\'enez Cruz}, N. M. and {Malhotra}, Ameek and {Tasinato}, Gianmassimo and {Zavala}, Ivonne},
  title   = {{Astrometry meets pulsar timing arrays: Synergies for gravitational wave detection}},
  journal = {Phys. Rev. D},
  volume  = {112},
  number  = {8},
  pages   = {083558},
  year    = {2025},
  doi     = {10.1103/8k1p-pzcg},
  eprint  = {2412.14010},
  archivePrefix = {arXiv},
  primaryClass  = {astro-ph.CO},
}

@article{Jaraba2025,
  author  = {{Jaraba}, Santiago and {Kuroyanagi}, Sachiko and {Liang}, Qiuyue and {Lin}, Meng-Xiang and {Trodden}, Mark},
  title   = {{First astrometric constraints on parity-violation in the gravitational wave background}},
  journal = {JCAP},
  volume  = {2025},
  number  = {08},
  pages   = {057},
  year    = {2025},
  doi     = {10.1088/1475-7516/2025/08/057},
  eprint  = {2505.18085},
  archivePrefix = {arXiv},
  primaryClass  = {astro-ph.CO},
}

@article{Liang2024,
  author  = {{Liang}, Qiuyue and {Lin}, Meng-Xiang and {Trodden}, Mark and {Wong}, Sam S. C.},
  title   = {{Probing parity violation in the stochastic gravitational wave background with astrometry}},
  journal = {Phys. Rev. D},
  volume  = {109},
  number  = {8},
  pages   = {083028},
  year    = {2024},
  doi     = {10.1103/PhysRevD.109.083028},
  eprint  = {2309.16666},
  archivePrefix = {arXiv},
  primaryClass  = {astro-ph.CO},
}

@article{Filipello2026,
  author  = {{Filipello}, Lorenzo and {Beordo}, William and {Akhmetov}, Volodymyr and {Crosta}, Mariateresa and {Lattanzi}, Mario Gilberto},
  title   = {{Combining astrometry with pulsar timing: the first joint analysis of very low frequency gravitational waves}},
  journal = {arXiv e-prints},
  year    = {2026},
  eprint  = {2609.21606},
  archivePrefix = {arXiv},
  primaryClass  = {astro-ph.CO},
}

@article{Moore2017,
  author  = {{Moore}, Christopher J. and {Mihaylov}, Deyan P. and {Lasenby}, Anthony and {Gilmore}, Gerard},
  title   = {{Astrometric Search Method for Individually Resolvable Gravitational Wave Sources with Gaia}},
  journal = {Phys. Rev. Lett.},
  volume  = {119},
  number  = {26},
  pages   = {261102},
  year    = {2017},
  doi     = {10.1103/PhysRevLett.119.261102},
  eprint  = {1707.06239},
  archivePrefix = {arXiv},
  primaryClass  = {astro-ph.IM},
}

@article{GolatContaldi2022,
  author  = {{Golat}, Sebastian and {Contaldi}, Carlo R.},
  title   = {{All-sky analysis of astrochronometric signals induced by gravitational waves}},
  journal = {Phys. Rev. D},
  volume  = {105},
  number  = {6},
  pages   = {063502},
  year    = {2022},
  doi     = {10.1103/PhysRevD.105.063502},
  eprint  = {2201.03903},
  archivePrefix = {arXiv},
  primaryClass  = {gr-qc},
}

@article{GaiaDR3,
  author  = {{Gaia Collaboration} and {Vallenari}, A. and others},
  title   = {{Gaia Data Release 3: Summary of the content and survey properties}},
  journal = {Astron. Astrophys.},
  volume  = {674},
  pages   = {A1},
  year    = {2023},
  doi     = {10.1051/0004-6361/202243940},
  eprint  = {2208.00211},
  archivePrefix = {arXiv},
  primaryClass  = {astro-ph.GA},
}

\end{document}